\documentstyle[spie]{article} 

\input{psfig.sty}   

\title{Capabilities of the NASA/IPAC Extragalactic Database in the Era 
of a Global Virtual Observatory} 

\author
{Joseph M. Mazzarella, Barry F. Madore, George Helou, 
and the NED Team\supit{a} 
\skiplinehalf 
California Institute of Technology, Jet Propulsion Laboratory, 
MS 100-22, Pasadena, CA 91125\\
\\
To appear in {\it Proceedings of SPIE: Astronomical Data Analysis},\\
eds. J. Starck \& F. D. Murtagh, Volume 4477, 2001 (in press).
}

\authorinfo{\supit{a}For additional information send correspondence to J.M.M. 
(mazz@ipac.caltech.edu) or B.F.M. (barry@ipac.caltech.edu).}
 
\begin{document} 
\maketitle 

\begin{abstract}

We review the capabilities of the NASA/IPAC Extragalactic Database
(NED, {\tt http://ned.ipac.caltech.edu}) for information retrieval and
knowledge discovery in the context of a globally distributed virtual
observatory. Since its inception in 1990, NED has provided astronomers
world-wide with the results of a systematic cross-correlation of
catalogs covering all wavelengths, along with thousands of extragalactic
observations culled from published journal articles.  NED is continuously
being expanded and revised to include new catalogs and published
observations, each undergoing a process of cross-identification to
capture the current state of knowledge about extragalactic sources
in a panchromatic fashion. In addition to assimilating data from
the literature, the team is incrementally folding in millions of
observations from new large-scale sky surveys such as 2MASS, NVSS, APM,
and SDSS. At the time of writing the system contains over 3.3 million
unique objects with 4.2 million cross-identifications.  We summarize
the recent evolution of NED from its initial emphasis on object name-,
position-, and literature-based queries into a research environment
that also assists statistical data exploration and discovery using large
samples of objects. Newer capabilities enable intelligent \lq\lq Web mining\rq\rq\
of entries in geographically distributed astronomical archives that
are indexed by object names and positions in NED, sample building using
constraints on redshifts, object types and other parameters, as well as
image and spectral archives for targeted or serendipitous discoveries.
A pilot study demonstrates how NED is being used in conjunction with
linked survey archives to characterize the properties of galaxy classes
to form a training set for machine learning algorithms; an initial goal
is production of statistical likelihoods that newly discovered sources
belong to known classes, represent statistical outliers, or candidates
for fundamentally new types of objects. Challenges and opportunities
for tighter integration of NED capabilities into data mining tools for
astronomy archives are also discussed.

\end{abstract}



\section{INTRODUCTION}
\label{sec:intro}  

The NASA/IPAC Extragalactic Database (NED) is widely acknowledged as the
most comprehensive and easy-to-use resource for information about sources
populating the Universe beyond our Milky Way galaxy\footnote[1]{E.g.,
{\it Nature}, Netguide, October 2000, also available online at
{\tt http://www.nature.com/netguide}.}. NED is an online research
facility designed to support scientists, educators, space missions and
observatories in the planning, execution and publication of research on
extragalactic objects. The foundation of NED is a growing database of
galaxies, quasars and all types of extragalactic objects that can be
searched by positions, redshifts, object types, references, authors,
and multi-wavelength cross-identifications produced from thousands of
catalogs and journal articles. The primary goal of NED is to maintain
an up-to-date panchromatic synthesis of basic data for all known
(cataloged and published) extragalactic objects, including pointers to the
astrophysical literature and to relevant distributed archive resources.
Scientists working in observational extragalactic astronomy use NED in
their research at nearly every step, from proposal planning, through
data collection, data interpretation, publication, and archiving of
calibrated images and spectra. Many professors also incorporate NED into
their lesson plans. As of the time of writing, over 2,100 articles in the
refereed astrophysics literature ({\it Astrophysical Journal, Astronomical
Journal, Monthly Notices of the Royal Astronomical Society, Astronomy \&
Astrophysics}, etc.) have acknowledged NED directly as an important tool
for the authors' research.

NED consists of these main components: (1) a growing database serving
cross-correlated panchromatic data and pointers for millions of
extragalactic objects; (2) a Web-based user interface accessible from
any Web browser with an Internet connection (as well as legacy VT-100
and X-window versions); (3) support for direct connectivity from remote
computer programs and Web sites that function as clients to the NED
servers; and (4) an automated batch mode for processing large requests.
In this paper we review highlights of each of these areas, outline
a case study currently under way that shows how NED is being used in
conjunction with linked survey archives to characterize the properties
of galaxy classes using machine learning algorithms, and conclude
with a discussion of how the capabilities will evolve in support of
large-scale data mining applications that can utilize NED in concert
with geographically distributed archives.  There is insufficient space
here for a comprehensive history and complete technical review of NED.
A discussion of the motivation, initial design, and early history of NED
was given by Helou {\it et al.}\cite{Helou91}.  Casual users, or those who
have not used it in a long time, may think of NED mainly as a `literature'
service or `digital library' in which users can only look up information
on objects one at a time by catalog entry name, or through article author
name searches and the like. In this paper we review the recent and ongoing
evolution of NED beyond these classic queries, including support for data
exploration and discovery using classes of objects.

\section{CROSS-IDENTIFICATIONS AND DATA INTEGRATION SERVICES}
\label{sec:xid}  

\subsection{Multi-wavelength Cross-Identifcations and Statistical Associations}

{\it Cross-identification} refers to the process of establishing
which observation in a specific catalog (for example, the FIRST radio
survey) corresponds to the same astrophysical source in surveys at other
wavelengths (for example, the far-infrared IRAS Faint Source Catalog).
The process is much more difficult that it may first appear, because
observations taken with different telescopes and at various wavelengths
often differ in substantial ways.  Positional uncertainties may differ by
factors of 2 to 10, or even more; and they may be ellipses with different
dimensions and position angles for each source (e.g., IRAS) rather than a
constant value across the sky. There are sometimes systematic errors in the
astrometry, such that positions in one survey will be offset from those in
another. If the angular resolution of one survey (e.g., X-rays from ROSAT)
is much lower than another (e.g., near-infrared detections from 2MASS),
more then one source in the second survey may contribute to the emission
seen in the first survey.  If only positional proximity is used to make
matches, different flux sensitivity limits and calibration uncertainties
(flux error bars) can lead to incorrect cross-identification between
a source in one catalog with a physically unrelated source in another
catalog that is nearby only in projection along the line of sight.  Also,
astrophysical sources can display very different structures at different
wavelengths because different physical processes are being observed.
For example, a radio source may have a core plus lobe emission located
on one or both sides of the core, but only the core (galaxy nucleus)
is typically detected in a survey at visual wavelengths. There is also a
strong wavelength dependency on the amount of extinction due to dust in the
interstellar medium of a galaxy; this can shift the centroid of a source
measured at a visual wavelength from that measured at an ultraviolet or
infrared wavelength.  In extreme cases like the merging system Arp 220,
a double central morphology in the blue band is produced by a dust lane,
while true double nuclei reside inside the dust lane and are visible
only at infrared and radio wavelengths.  Finally, objects populate
a hierarchical Universe: active nuclei, supernovae, and star-forming
regions reside in their host galaxies; many galaxies are members of pairs
and groups; galaxies, pairs and groups are typically members of clusters,
and galaxy clusters reside in superclusters separated by vast voids.

For these reasons, complex relationships between objects are needed (e.g.,
one-to-one, one-to-many, many-to-one, many-to-many), in addition to {\it
statistical associations} for cases in which simple, confident one-to-one
relationships cannot be established. NED activities revolve around a
systematic process of establishing realistic cross-identifications and
statistical associations between millions of entries in multi-wavelength
catalogs and publications. Cross-identifications are established in an
iterative way, being refined as new information becomes available. The
NED team works closely with the astronomical community in this process,
often resolving disputes and documenting errors in extensive notes
that are made available to users. It is sometimes assumed that the
NED team establishes cross-IDs in an `old-fashioned', manual way.
In fact, the process involves a fairly sophisticated computer program
that cross-correlates the positional uncertainties of positions in
NED with those in a new input catalog; the output is sorted into lists
containing (a) sources that are obviously new to NED, (b) a list of secure matches
(cross-identifications) with previously known sources, 
and (c) a list of \lq fuzzy\rq\rq\ cases
that need follow-up analysis of the statistical association parameters
for the many reasons outlined above.
The association parameters include 
the separation in arcseconds, the position angle in degrees, and
dimensionless parameters {\bf r} and {\bf p} that
represent measures of the \lq\lq goodness of fit\rq\rq\ 
of the convolved positional uncertainty ellipses of an
input object and a nearby NED object.\footnote[2]
{
The first parameter, {\bf r}, is the distance between the two sources in units of the
standard deviations of the convolved uncertainties on the principal axes
of the error ellipsoid; mathematically, this is the chi-square parameter
with two degrees of freedom evaluated for the observed separation and
catalog uncertainties, assuming Gaussian errors, and is dimensionless. 
The second parameter, {\bf p} (stored in NED as a base-10 logarithm), is the
expected-error density function evaluated for the observed separation. The
expected-error density function is the convolution of the error density
functions for the two catalogs involved, assuming Gaussian errors; the
density function has units of probability mass per steradian. 
More information about these parameters is available in NED's online documentation.
}
The NED interface currently makes these catalog source association
parameters available to users before they are worked off and turned
into cross-identification (where possible); prior to October 2000 these
parameters were utilized only internally by the NED team.

\subsection{Database Management, Growth and Related Activities}

NED does not simply ingest complete catalogs and maintain them in their
original form and format. While fundamental data such as positions,
redshifts, sizes, and flux/magnitude measurements are assimilated
into NED for the purposes of cross-identification and construction
of multi-wavelength SEDs, other data are connected in context using
pointers to remote archives. For example, extended sources in the Two
Micron All Sky Survey (2MASS) and the Sloan Digital Sky Survey (SDSS)
catalogs have many types of magnitude measurements. Only the magnitudes
from the recommended \lq\lq default\rq\rq\ method from each survey catalog
are folded directly into NED; the rest are being made easily accessible
using hyperlinks that query complete catalog entries at the survey/mission
archive sites. NED provides hyperlinks that issue queries of many major
archive services (e.g., {\it High Energy Astrophysics Research Center
[HEASARC]}, {\it Multi-Mission Archive at STScI [MAST]}, {\it Infrared Science Archive
[IRSA]} at IPAC, {\it SIMBAD} and {\it VizieR} services at {\it CDS} [Strasbourg, France],
{\it NRAO} and the {\it NASA Astrophysics Data System [ADS]} abstract service
and its linked journal Web sites) in a convenient \lq\lq 1-click\rq\rq\ fashion
that utilizes source names, survey/catalog cross-identifications, and sky
positions as the \lq\lq glue\rq\rq\ between the distributed data sets. An
illustration of this innovative capability (available online in NED since
April 2000) is given in Figure~\ref{fig:arc}.  The database contents and
relationship pointers are revised and augmented constantly to keep up
with new online survey data and knowledge appearing in the literature.
Updates to the public database occur approximately every three months
after periods of data entry, quality assurance checks, and testing using
an internal development version.

Important recent additions to NED's holdings include extragalactic
supernovae, the {\it Hubble Deep Fields (North \& South)}, the {\it
FAINT IMAGES OF THE RADIO SKY AT TWENTY-CENTIMETERS (FIRST), Eighth
Cambridge Radio Catalog (8C), Molonglo Reference Catalog
(MRC), Texas, Westerbork} and {\it MIT-Greenbelt} radio surveys, the {\it
Automated Plate Measurement (APM) Bright Galaxy Catalog}, the {\it Canadian
Network for Observational Cosmology (CNOC) Catalog}, and the {\it Las
Companas Redshift Survey (LCRS)}.  In addition to folding in data appearing
in the current literature, the team is establishing cross-identifications
and probabilistic associations between new observations from large surveys
(more than $10^6$ objects) and previously known sources in NED. Large
surveys which are being assimilated in an incremental fashion at the time
of writing include the {\it Two Micron All Sky Survey (2MASS), NRAO VLA
Sky Survey (NVSS), Automated Plate Measurement/United Kingdom Schmidt
(APM/UKS)}, and {\it Sloan Digital Sky Survey (SDSS)}.  The total data
holdings have been roughly doubling each year.  Larger database disks have
recently been configured to accommodate a system containing about ten
times the current holdings, enough for essential data and relationships
for about 50 million objects. Ongoing upgrades to the data management
and catalog cross-correlation and association processes, combined with
the rapid rates of increasing computer speed and decreasing data storage
costs, will allow NED to scale up its data integration activities to handle
order of magnitude increases in the number of unique extragalactic sources
(and candidates) with their cross-identifications and associations in
coming years.

NED staff works in coordination with other NASA archive centers, referred
to collectively as the Space Science Data System (SSDS),\footnote[3]{\tt
http://ssds.nasa.gov/} the CDS (Strasbourg, France)\footnote[4]{{\tt
http://cdsweb.u-strasbg.fr}}, the AAA publications board, Journal
editors, authors and referees, IAU Working Groups on Nomenclature, Data
and Journals, and the broader astronomical community to improve data
handling and archive services.  The team provides extensive user support
(Help Desk) to answer questions and take users' input for priorities
on new developments.\footnote[5]{Inquiries may be emailed to {\tt
ned@ipac.caltech.edu}.}

\subsection{DATABASE CONTENTS}

\begin{figure}[!htb]
   \begin{center}
   \begin{tabular}{c}
   \psfig{figure=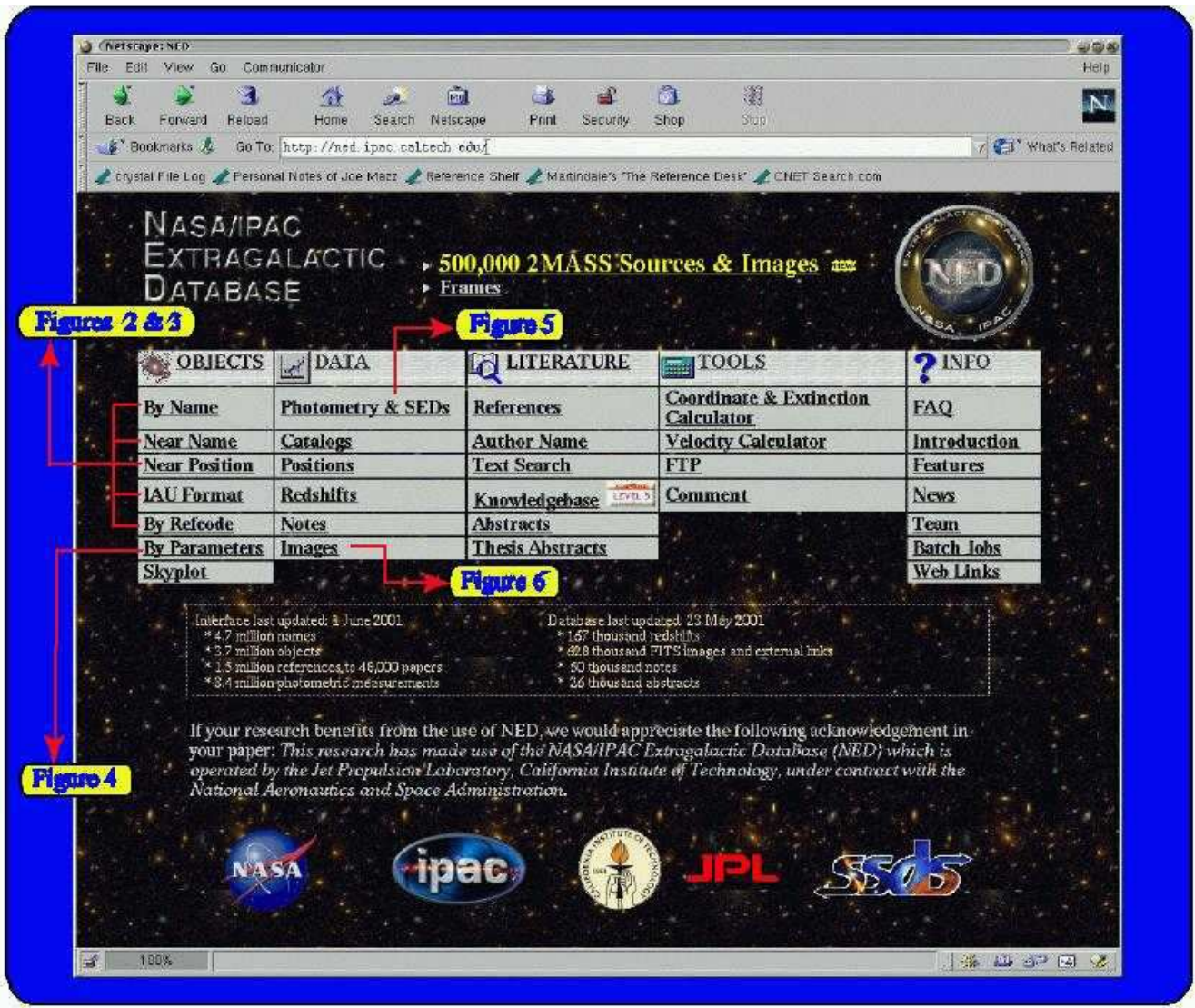,width=6.9in,angle=0} 
   \end{tabular}
   \end{center}
   \caption[fig1] 
   { \label{fig:home}	  
The NED Web interface main menu available at {\tt http://ned.ipac.caltech.edu}.
The annotations refer to example query results which are displayed in 
Figures 2-6.} 
   \end{figure}

The database contents of NED at the time of 
writing (July 2001) are summarized in Table~\ref{tab:contents}.
This information is updated periodically on the NED home page 
(Figure~\ref{fig:home}) after each update to the public database.

\begin{table} [h]   
\caption{NED database contents as of July 2001.} 
\label{tab:contents}
\begin{center}       
\begin{tabular}{|l|}
\hline
\rule[-1ex]{0pt}{3.5ex} 
4.7 million cross-identifications in thousands of multi-wavelength surveys and journal articles\\
\hline
\rule[-1ex]{0pt}{3.5ex} 
3.7 million unique extragalactic objects \\
\hline
\rule[-1ex]{0pt}{3.5ex} 
3.4 million photometric measurements covering gamma-rays through radio 
wavelengths\\ 
\ \ (with uncertainties) and dynamic spectral energy distribution plots  \\
\hline
\rule[-1ex]{0pt}{3.5ex} 
2.0 million detailed position measurements with uncertainties  \\
\hline
\rule[-1ex]{0pt}{3.5ex} 
1.5 million bibliographic references to 48,000 articles \\
\hline
\rule[-1ex]{0pt}{3.5ex} 
167,000 redshifts from the published literature  \\
\hline
\rule[-1ex]{0pt}{3.5ex} 
628,000 science-grade FITS images and remote links with \lq\lq thumbnail\rq\rq\ previews \\
\hline
\rule[-1ex]{0pt}{3.5ex} 
50,000 detailed notes from catalogs and journal publications  \\
\hline
\rule[-1ex]{0pt}{3.5ex} 
26,000 journal article abstracts and 1,150 Ph.D. thesis abstracts \\
\hline
\end{tabular}
\end{center}
\end{table}

The essential data for sources in NED include positions, redshifts,
morphological types, nuclear spectral types, panchromatic photometry, and
images. When available, uncertainties in the measurements are also stored
and provided by the interface.  Photometry data are stored in original
units and converted to common frequency (Hz) units and flux density units
($W~m^{-2}~Hz^{-1}$) for construction of Spectral Energy Distributions
(see Figure~\ref{fig:seds}); the data are also tagged with their aperture sizes or
status as a \lq\lq total flux\rq\rq\ measurement. The extragalactic
objects types available in NED are summarized in Table~\ref{tab:classes}.

\begin{table} [!htb]   
\caption{Extragalactic object types in NED.} 
\label{tab:classes}
\begin{center}       
\begin{tabular}{|l|l|l|} 
\hline
\rule[-1ex]{0pt}{3.5ex} Galaxies & QSOs & Radio Sources  \\
\hline
\rule[-1ex]{0pt}{3.5ex} Galaxy Pairs & QSO Groups & Infrared Sources \\
\hline
\rule[-1ex]{0pt}{3.5ex} Galaxy Triples & Gravitational Lenses & Visual Sources \\
\hline
\rule[-1ex]{0pt}{3.5ex} Galaxy Groups & Absorption Line Systems & UV Excess Sources \\
\hline
\rule[-1ex]{0pt}{3.5ex} Galaxy Clusters & Emission Line Sources & X-ray Sources \\
\hline
\rule[-1ex]{0pt}{3.5ex}  & Supernovae & Gamma-ray Sources \\
\hline
\end{tabular}
\end{center}
\end{table}

   \begin{figure}[!htb]
   \begin{center}
   \begin{tabular}{c}
   \psfig{figure=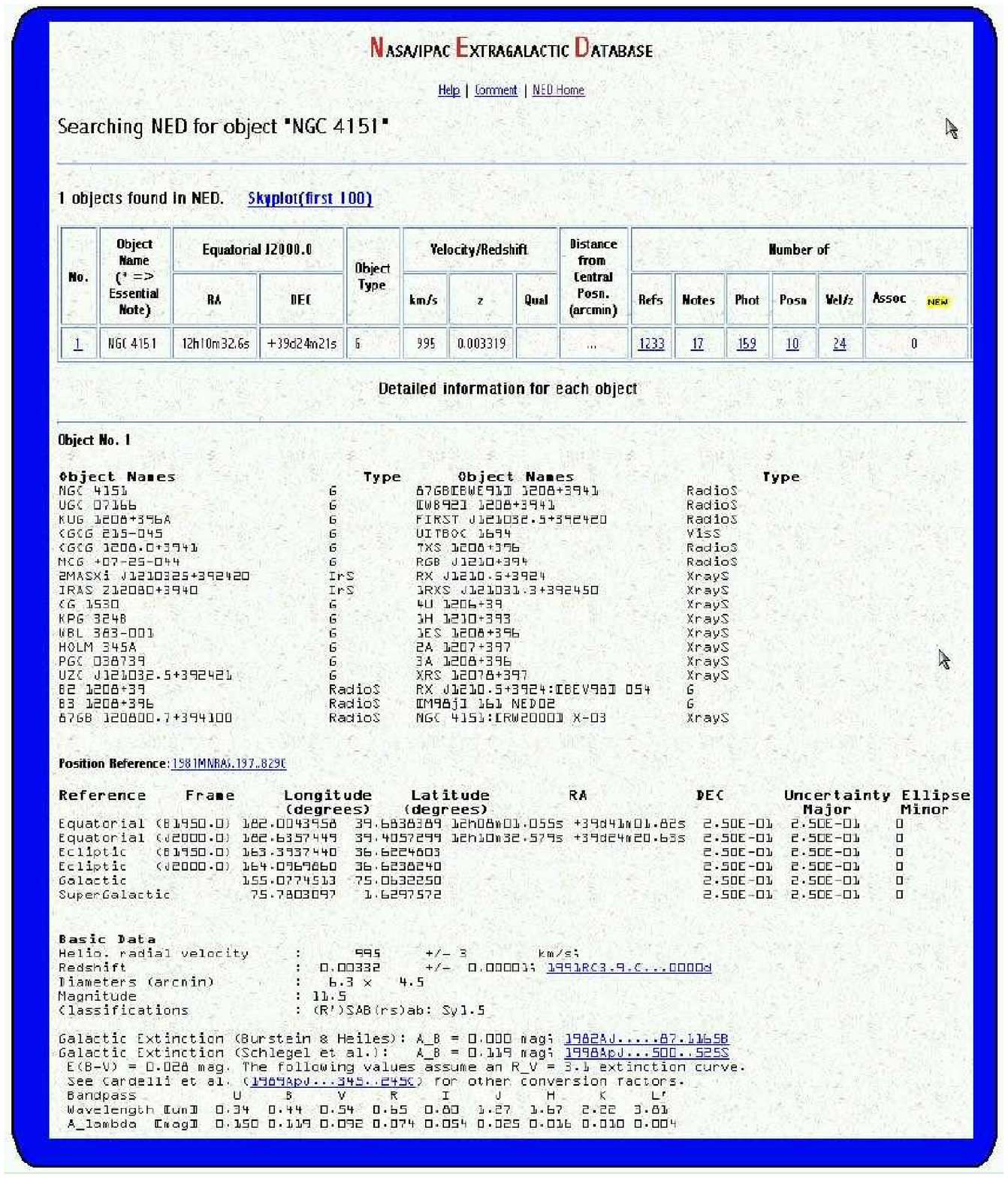,height=5.5in,angle=0} 
   \end{tabular}
   \end{center}
   \caption[fig2] 
   { \label{fig:data}	  
Essential Data returned by NED from a query of NGC 4151: includes 
coordinates and redshift (with uncertainties), 
multi-wavelength survey cross-identifications and object types, 
size, magnitude, classifications, 
Galactic extinction along the line of site, as well as links to 
query references, notes, photometry, positions, velocities, and images.
This information is followed by links to {\bf External Archives} (Figure~\ref{fig:arc}).} 
   \end{figure} 

\section{INTERFACES AND QUERY SERVICES} 
\label{sec:int}


NED is accessible on the World Wide Web at {\tt
http://ned.ipac.caltech.edu}. The Web interface is by far the most common
way that most users access NED; usage has grown from about $10^5$ requests
per month in 1997 to an average of over $10^6$ requests per month in 2001.
The rate of growth of access to the NED Web server is also rising quickly. 
There is also a VT-100 (ASCII) and X-Window graphical user interface
available via telnet login session (`{\bf telnet ned.ipac.caltech.edu}').
These legacy services provide access to the same database updates as
available through the Web interface, but they are no longer maintained
or enhanced with new services.  Limited resources to build new tools are
going into the much more heavily used Web and batch access modes.

\subsection{Web Query Services}

Figure~\ref{fig:home} shows the NED Web interface main menu (home page),
where the main functionality is organized in columns as follows.

\subsubsection{Objects}

In the {\it OBJECTS} column NED can be searched for extragalactic objects
using menus designed for searches `By Name', `Near Name' (input a name
and search radius), `Near Position' (input celestial coordinates and
a search radius), `IAU Format' (object names), or `By Reference Code'
(a 19 digit code developed jointly by NED with the ADS\footnote[7]{\tt
http://adswww.harvard.edu} and CDS to uniquely identify publications in
astronomy).  The `Skyplot' service generates finder charts indicating
NED objects and star positions in a specified field of view.
Figure~\ref{fig:data} illustrates an example of the Essential Data that
is available from NED after a query of the galaxy NGC 4151.

\begin{figure}[!htb]
   \begin{center}
   \begin{tabular}{c}
   \psfig{figure=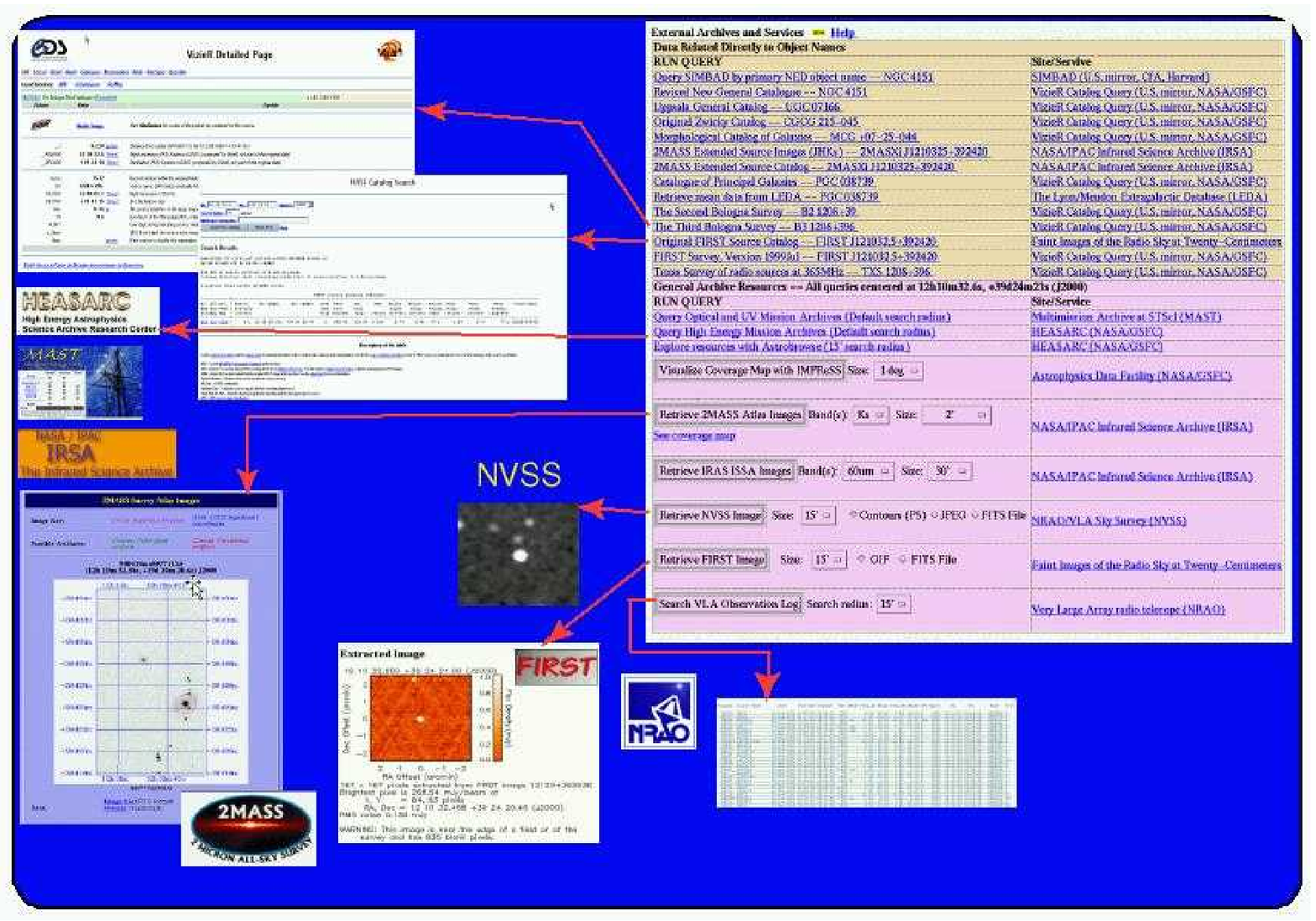,width=6.9in,angle=0} 
   \end{tabular}
   \end{center}
   \caption[fig3] 
   { \label{fig:arc}	  
Queries to globally distributed archive data are conveniently available with a
single mouse click.  Such links are generated dynamically during 
object query report generation.  
The first set of hyperlinks are to data related
directly to an object name; the second set of hyperlinks are to archive 
services that can be queried at the coordinates of the NED object.
This example is a continuation and extension of the report from a
query of NGC 4151 (Figure~\ref{fig:data}). 
}
   \end{figure}

\subsubsection{Global Archive Connectivity}
\label{sec:con}


Figure~\ref{fig:arc} shows hyperlinks to `External Archives and
Services', easily found when the user scrolls down below the Essential
Data that comes directly from the NED database. Links here allow the
user to retrieve images and query original catalog data or observation
log entries.  The first set of hyperlinks are to data related directly
to an object name; the second set of links are to services that can
be queried at the object's coordinates. A summary of the available
resources includes: original catalog record entries in VizieR at
CDS/France; 
the NVSS, and FIRST catalog and image servers and the 
Observation Log of the VLA telescope from the 
{\it National Radio Astronomy Observatories (NRAO)}; 
infrared mission archives at {\it IRSA/IPAC} (2MASS, IRAS, etc.); 
visual and UV mission archives at {\it MAST/STScI} (HST, IUE, etc.); 
high energy mission archives at {\it HEASARC/GSFC}
(ASCA, CGRO, Einstein, etc.); 
and the IMPRESS focal plane plotting service for NASA 
astrophysics missions at the {\it NASA Goddard Space
Flight Center/Astrophysics Data Facility 
(GSFC/ADF)}\footnote[8]{
\tt http://cdsweb.u-strasbg.fr, http://www.nrao.edu,
http://irsa.ipac.caltech.edu,
http://archive.stsci.edu,
http://heasarc.gsfc.nasa.gov,
http://adf.gsfc.nasa.gov/adf/adf.html
}.  
This tool makes it very easy for
researchers to determine whether an object has been observed by one or
more of the major surveys or observatories. New archive services are being added
as they become available. With this service, users now have at their
fingertips distributed data dynamically cross-linked (federated, fused)
using source names and positions indexed and maintained by NED.  {\it This
innovative service is a major step toward federating distributed archives
as discussed in reference to a National Virtual Observatory portal.}

   \begin{figure}[!hp]
   \begin{center}
   \begin{tabular}{c}
   \psfig{figure=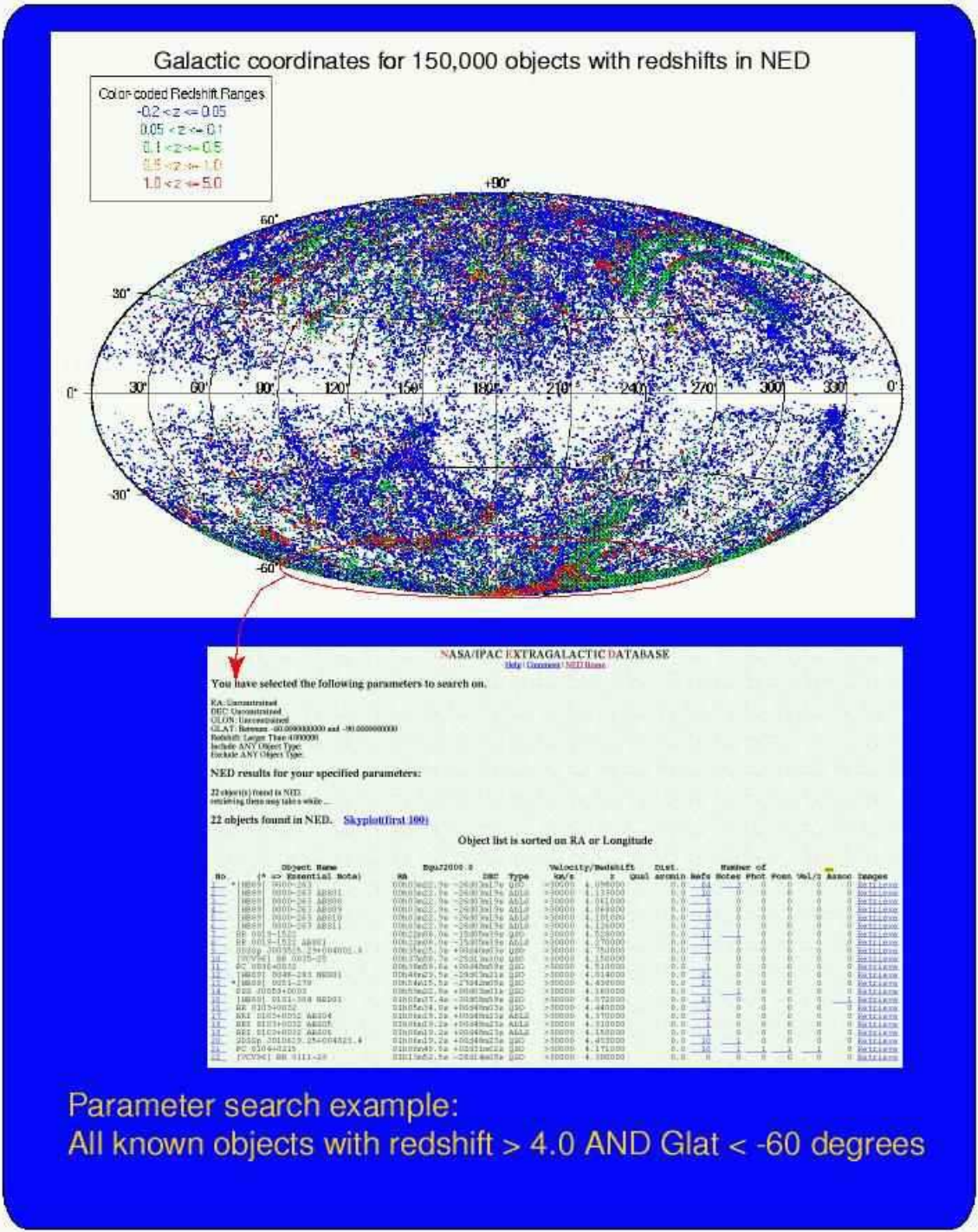,width=6.0in,angle=0} 
   \end{tabular}
   \end{center}
   \caption[fig4] 
   { \label{fig:byparam}	  
Objects that meet the search criteria 
[{$Redshift > 4.0\ AND\ Glat < -60^{\circ}$}] returned after
submitting a query using the `By Parameters' menu (see Figure 1).
} 
   \end{figure} 

\subsubsection{Sample Building with `By Parameter' Queries}

The `By Parameters' menu allow the user to query NED using joint
constraints on redshift range, sky area, object types, or survey/catalog
name prefixes.  Casual users of NED may not realize the benefits of
querying NED using a catalog name prefix. With this feature one can
dynamically regenerate a classic catalog sample that contains not simply
the original catalog measurements (available elsewhere, such as VizieR),
but rather the most precise and currently available source positions
and redshifts, with links to up-to-date references, multi-wavelength
photometry, images, etc. For example, today anyone can use NED to generate
an up-to-date compilation analogous to the `{\it Catalog of Markarian
Galaxies}'\cite{Mazz86}, or generate a current data set for the entire
{\it Third Cambridge (3C)} radio galaxy sample, with the click of a mouse.
Since entries in NED are continuously updated through a synthesis of the
literature and large surveys, errors in original catalogs are often found
and documented. Therefore, {\it for many studies it is more efficient and
effective to cross-correlate new observations against the data synthesis
in NED rather than against catalogs in their original published forms}.
A visualization of one example of this powerful feature is shown in
Figure~\ref{fig:byparam}. Other example queries that can be performed with
the `By Parameters' tool are (including the number of objects returned
at the time of writing):

\begin{itemize}
\item Find known objects with a redshift greater than 3.0: 1034 objects
\item Find known objects with a redshift greater than 3.0 which are 
  not QSOs or absorption line systems: 154 objects
\item Find extragalactic radio sources (from ANY catalog/survey) 
  with a redshift less than 0.05 and Galactic latitude greater than 
  $50^{\circ}$: 735 objects
\item Report up-to-date data and pointers for all extragalactic objects in the 
8C radio sample: over 6200 objects
\end{itemize}

Planned upgrades to this functionality are reviewed below (Section~\ref{subsec:future}).

\subsubsection{Data}

   \begin{figure}[!htb]
   \begin{center}
   \begin{tabular}{c}
   \psfig{figure=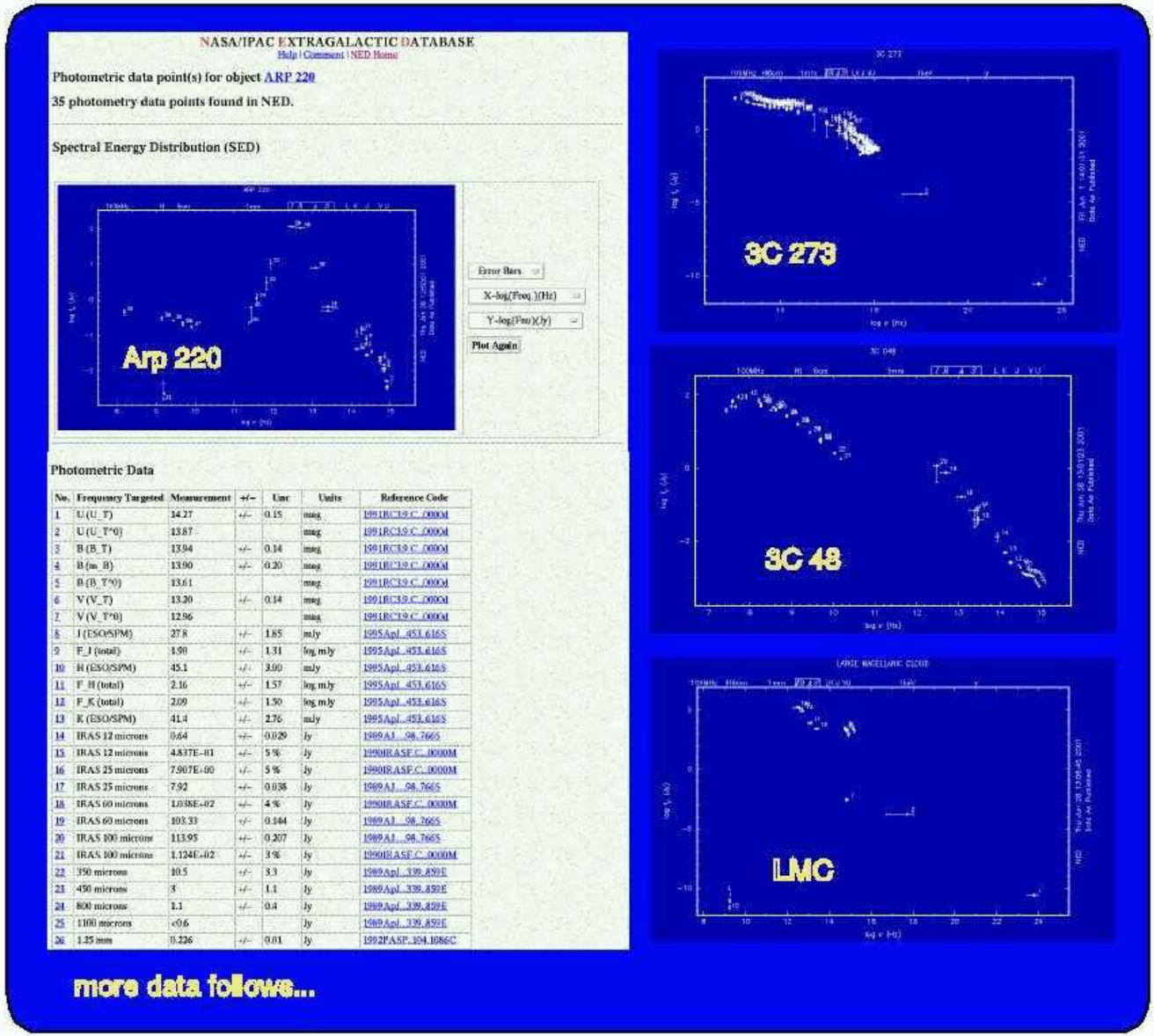,width=6.9in,angle=0} 
   \end{tabular}
   \end{center}
   \caption[fig5] 
   { \label{fig:seds}	  
Multi-wavelength photometry and spectral energy distributions (SEDs) covering
gamma-ray through radio wavelengths.
The data are available in original units as published, and also 
in common units ($Hz, W m^{-2} Hz^{-1}$) used to construct
SED plots. The data include uncertainties and references. 
The SED plots are dynamic, with configurable axis units 
(e.g., wavelength or frequency for the abscissa and
$f_{\nu}$, $\nu f_{\nu}$ or $f_{\lambda}$ for the ordinate) 
and optional error bars.
} 
   \end{figure} 

The {\it DATA} column in the main NED menu allows the user to enter an object name
(e.g., `NGC 4151', `APMUKS(BJ) B003425.77-334949.0', `SN 1993G',
`2MASXi J1132350+582422', `SDSS J1044-0125', `Antennae') and query NED for
`Photometry \& SEDs', `Catalogs', `Positions', `Redshifts', `Notes', 
or `Images. In the report resulting from a query of any NED object,
hyperlinks are included that provide access to this same information;
the {\it DATA} search menus simply provide more direct access.
Figure 5 illustrates an example of multi-wavelength photometry data
and Spectral Energy Distribution (SED) plots available from NED.

\begin{figure}[!htb]
   \begin{center}
   \begin{tabular}{c}
   \psfig{figure=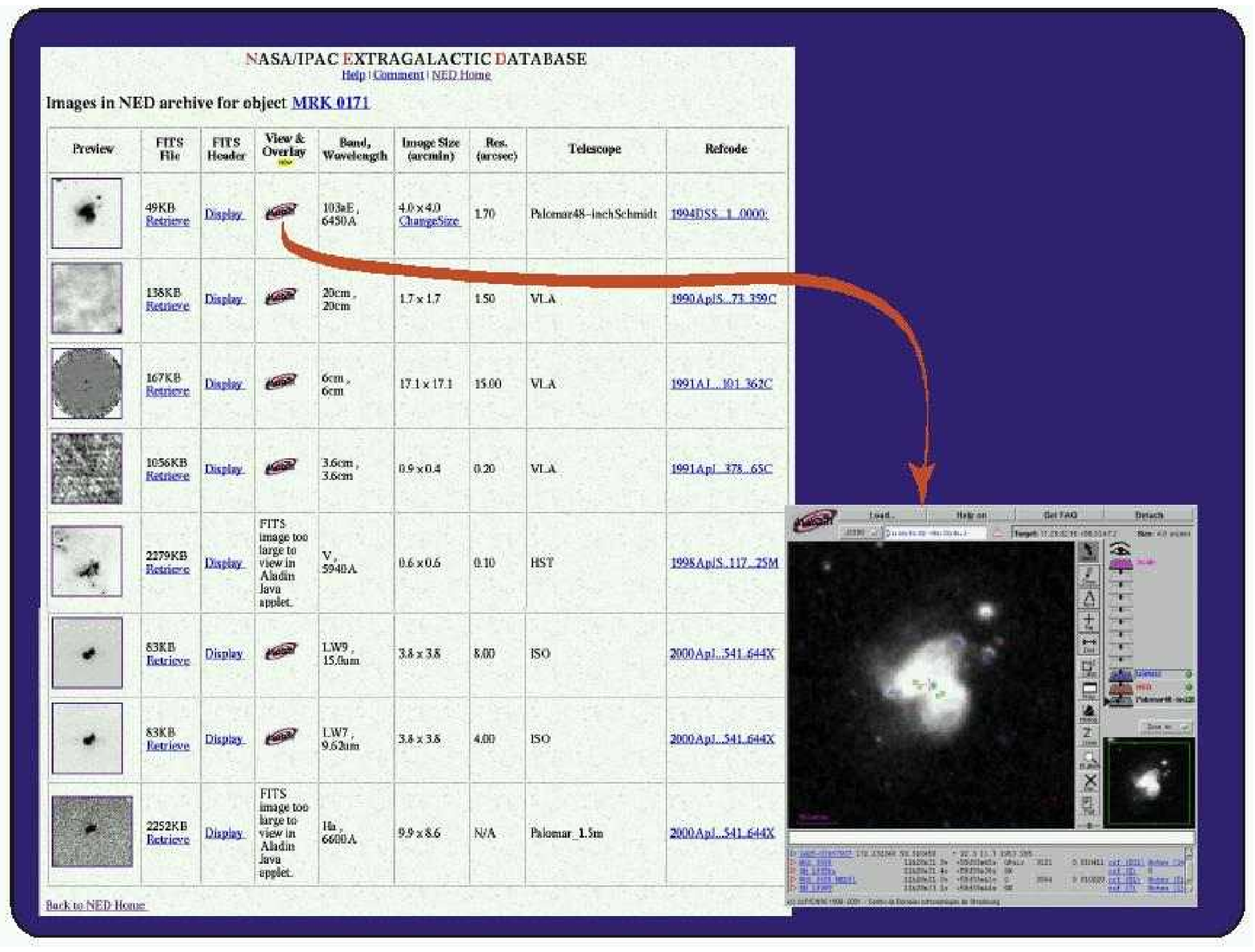,width=6.9in,angle=0} 
   \end{tabular}
   \end{center}
   \caption[fig6] 
   { \label{fig:images}	  
Multi-wavelength galaxy images,
including previews (GIF) and science-grade data (FITS).
Image overlays and graphical inter-activity
between sky coordinates (from information in the FITS image header) and object
markers (from NED and other databases) are available using the 
Aladin Java applet (CDS).
} 
   \end{figure}

Figure~\ref{fig:images} shows an example of the rich variety of
multi-wavelength FITS images available for immediate download or
visualization using the Aladin Java applet.  With Aladin the user
can: superimpose entries from NED, the USNO catalog and various other
astronomical catalogs; interactively link to information for all known
objects in the field; measure interactively positions and distances
in celestial coordinates.  Aladin's interoperability with NED and other
distributed data services provides a visual summary of the multi-wavelength
sky. Aladin was developed at the CDS (Strasbourg, France) and is configured with the
NED interface through a cooperative agreement.  The NED team has made
significant contributions to testing and debugging of Aladin using the
richness of FITS image types in the NED archive.  This is a successful
example of software tool reuse to meet common goals and interests through
an international collaboration.

\subsubsection{Literature}

In the {\it LITERATURE} column of the main NED menu, the user can:
(1) enter an object name and access the `References' related to that
object; (2) enter an `Author Name' and retrieve a list of references;
(3) search journal article `Abstracts' by year, volume or page numbers;
(4) search `Thesis Abstracts' by year range;
(5) use the `Text Search' tool to perform a keyword search on the indexed
contents of the journal and thesis abstracts in NED or the full text
content of LEVEL5; and (6) access the LEVEL5 `Knowledgebase'.

LEVEL5 can also be accessed directly
at {\tt http://ned.ipac.caltech.edu/level5}. It contains
hyperlinked review articles and documents of current and lasting interest
to cosmologists and extragalactic astronomers.  Contents include a
glossary of terms, essays, recent research articles, detailed monographs
and extensive reviews (where copyrights allow).  Within each article,
cited extragalactic objects are cross-linked to NED Basic Data frames,
and all available citations are hyperlinked to NASA's Abstract Data Service
(ADS), to on-line NED abstracts, or to preprints on astro-ph. Tabular data,
images and graphs are being progressively linked to and from relevant
essays and review articles. 

\subsubsection{Tools}

The {\it TOOLS} column of the main NED menu contains a `Coordinate \&
Extinction Calculator' that performs coordinate conversions and precession,
and displays line-of-sight Galactic extinction estimates using two modern
techniques\cite{Sch98,Bur82}. The `Velocity Calculator' computes
conversions between redshifts for extragalactic objects in different
reference frames: heliocentric, Local Group, Galactic Standard of Rest,
and 3K Microwave Background. There is also a link to NED's public
FTP site, primarily used by users to pick up their output from the 
NED batch mode.

\subsubsection{Information}

The {\it INFO} column of the main menu contains links to a Frequently Asked Questions
(`FAQ') document, an `Introduction', summary of `Features', `News',
information about the NED development `Team', forms and documentation for submitting
NED `Batch jobs', and finally a page of useful `Web Links' relevant
to extragalactic astronomy.

\subsection{Batch Mode}

NED can process requests for large amounts of data through its `Batch
Job' option. Using this mode simply involves submitting to NED via email
a \lq\lq batch form\rq\rq\ containing a list of objects or positions,
or other constraint parameters (e.g., redshift, object type, or survey/catalog
name prefix).\footnote[5]{
The forms are available from the `Batch Jobs'
link in the {\it INFO} column on the main NED menu. 
The forms can be emailed to {\tt nedbatch@ipac.caltech.edu} for processing.}
After the request has been processed, NED sends the user
a notice by return email indicating where the resulting data files may
be copied via FTP.  There are two types of batch job forms available. One
form is to search any of the main data categories in NED -- Objects, Basic
Data, References, Photometry, Positions, and Redshifts; the second form
is to constrain searches By Parameters -- positions, names, object types,
and redshifts. The input forms are flexible enough to support several
different searches with a single form. Though the batch processor will
currently support only 3,000 input requests per job, it will return up
to 10,000 objects per request.  See Section~\ref{subsec:future} for a
summary of planned upgrades to the batch mode.

\subsection{Client/Server Mode Connectivity}
\label{subsec:server}

For many years NED has provided a `server mode' with custom client
(C) software \footnote[6]{The NED client C code is available at {\tt
ftp://ned.ipac.caltech.edu/pub/ned/client.2/}.} that has been used
by computer programmers all over the world to build applications that
issue queries and retrieve data from the NED database in a format that
can be integrated into their services. Astrophysics archive centers and
observatories use NED's server mode extensively to resolve extragalactic
object names into celestial coordinates, and to retrieve lists of
objects by specifying an input position and search radius. A number
of sky visualization tools also make use of NED's client/server mode,
including IPAC's {\it IRSKY} and {\it SIRTF Planning Observations Tool
(SPOT)}\footnote[7]{For information about Skyview and SPOT see the IPAC
Web site at {\tt http://www.ipac.caltech.edu}.}, and the {\it Scientist's
Expert Assistant (SEA)} developed at GSFC and STScI for HST and NGST
proposal planning\footnote[8]{{\tt http://aaaprod.gsfc.nasa.gov/sea}}. The
NED client/server service is being updated with a new approach 
as discussed below (Section~\ref{subsec:xml}).

\subsection{Current and Future Technologies}
\label{subsec:xml}

NED is currently operated using UNIX (Sun/Solaris) servers, a combination
of Sun D-1000 disk arrays and miscellaneous SCSI disk \lq\lq shoe
boxes\rq\rq\ (totaling over 650 GB, including a backup system), two
Informix relational database servers, Apache Web servers, and custom
Web interface software (CGI) written mostly in C.  Perl is also used
for some applications, as well as Java applets.  New technologies
are evaluated as time allows, and applied only when there is a clear
benefit in capabilities or performance over current approaches. As for
most operational systems with a large base of users, limited resources
require a gradual transition to new technologies, because legacy services
have to be maintained during any transition. A relatively new technology
planned for future NED interface enhancements is a Web application server
(`middle-ware') using Java Servlets and Enterprise Java Beans (EJBs); this
will enable a more sophisticated interface that includes customized options
stored across user sessions, load balancing across multiple servers,
and better performance and scalability compared to current CGI-based
Web services. The NED team, in coordination with the CDS, IRSA, and the
ADC/ADF group at GSFC, is also designing new formats for data output from
the NED Web (HTTP) servers based on eXtensible Markup Language (XML).
XML is a markup language designed for producing logical, well structured
documents and data services that separate data content and metadata issues
from those of display and presentation\footnote[9]{For information about
XML see {\tt http://www.xml.org}, {\tt http://www.w3.org/XML}, and the
astronomical XML resources at {\tt http://xml.gsfc.nasa.gov}. There
are dozens of books published on the subject.}.  This development will
simplify the ability of client computer programs to be written that can
parse and utilize NED data in more efficient ways than is possible using
the classic NED `server mode' (described in Section~\ref{subsec:server}),
or by parsing the HTML output which is expected to undergo frequent changes
to provide improved presentation for NED's interactive users. Providing
XML output from NED will enable, among other things, the construction
of software \lq\lq agents\rq\rq\ that perform automated data mining
by streaming and combining data from NED and other services, improved
interoperability with other services that are planned by developers of
new Virtual Observatory applications, and support for the next generation
of Web browsers and data analysis software that will increasingly utilize
XML rather than HTML.  Providing XML output from NED will establish
the next generation `server mode`, superseding the need to enhance the
custom NED client/server package (Section~\ref{subsec:server}), because
builders of future observatory and archive center tools will find it
more convenient and powerful to connect their software with NED using
industry-standard XML development tools.

\section{NED IN THE ERA OF A GLOBAL VIRTUAL OBSERVATORY}
\label{subsec:vo}




In a nutshell, the basic vision of the `virtual observatories' (VO)
involves interconnected, globally distributed archives from observatories
and large-scale sky surveys which are federated using common database
query standards and data interchange protocols, combined with user
interfaces and data mining tools that integrate and analyze the fused,
multi-wavelength data sets to facilitate making new discoveries about the
Universe (regardless of the location of the data or the investigators). A
popular level description of the concept is given in an article by Cowen
\cite{SciNews2001}. The term `virtual observatory' is new, but the concept
is essentially the same as that outlined a decade ago for the original
Astrophysics Data System (ADS)\cite{Weiss91}. Portions of the VO concept
also overlap, to some degree, the functionality of NED with respect
to the production of multi-wavelength catalog cross-identifications
and statistical associations, queries that join distributed archives,
and support for studies of fused panchromatic data for extragalactic
sources. It is noteworthy that Helou {\it et al.}\cite{Helou91} pointed
out in 1991 that there is a dual challenge presented by the explosive
growth in astronomical data: \lq\lq dealing with the sheer volume,
but also inter-connecting intelligently the huge variety of information
available.\rq\rq\ Yet there is much work left to be done on all fronts
by a broad community of astronomers teamed with computer scientists and
programmers. The NED team shares a common vision regarding what a VO can
enable for all fields of astrophysics, and we are actively involved in
numerous collaborations and proposals designed to lay the foundation of
the VO and extend its functionality. It is useful to summarize the role
of NED in the emerging global VO in the context of current capabilities
and logical future enhancements that will support related projects.

\subsection{Current Capabilities}

As the primary astrophysics data integration service of NASA's
Space Science Data System\footnote[1]{\tt http://ssds.nasa.gov}, in
cooperation with the Astrophysics Data Center Coordinating Council
(ADCCC)\footnote[2]{\tt http://www.adccc.org} and new VO partners,
NED will continue to establish and improve high fidelity relationships
between multi-wavelength data with anchors to the literature for millions
of extragalactic objects, using a combination of computer software that
utilizes positional uncertainties and astronomical source properties (e.g.,
fluxes, redshifts, sample source densities), followed by human inspection
to resolve important, complex cases that cannot be fully automated. In
addition, through collaboration on a number of proposals to NASA and the
National Science Foundation, the NED team is committed to participation
in the collaborative development, testing, and deployment of the next
generation of catalog cross-identification software tool-kits. The
idea is that new tools that may be developed for individual researchers
to construct dynamic catalog cross-identifications in a VO framework,
for example enabling Bayesian type probabilities of association that
incorporate prior astrophysical knowledge about a sample\cite{Rut2000},
will also be useful for performing rapid, bulk cross-identifications of
new surveys against the data synthesis in NED. The advantages of this
approach (over repeatedly performing dynamic cross-comparisons against
a plethora of original catalogs) for many types of studies were reviewed
in Section~\ref{sec:xid}.

NED will also continue to maintain its object-based portal into
distributed data sets (Section~\ref{sec:con}, Figure~\ref{fig:arc}),
in order to make this innovative service even more useful for a wide
cross-section of the extragalactic research community. This work will
involve keeping up with evolving technology for interoperability between
archive query services, primarily XML and its associated family of protocols
(e.g., XLink, Namespaces, DTD, Schema, CSS, XSL, SOAP) \footnote[3]{See
{\tt http://www.w3.org} for descriptions of these and other XML protocols
put forth by the World Wide Web Consortium.} as they are adopted by the
community of VO developers (along with the much broader Internet industry).

\subsection{Future Enhancements}
\label{subsec:future}

In addition to staying current with the literature and extragalactic
source observations in modern large-scale sky surveys, the NED team
is committed to providing new functionality to extend the usefulness
of NED as a research tool for astronomers. The newer NED capabilities
with direct relevance to the VO concept were reviewed above.  Over the
next few years NED will provide the following enhanced capabilities
for even tighter integration with other VO initiatives: (1) development
of a spectral archive for extragalactic sources, focusing on reduced,
calibrated spectra contributed by authors and links to spectra in
mission/observatory archives (very similar to the NED image archive);
(2) enhancements to the `By Parameters' tool to support all-sky queries
based on multi-wavelength flux and color criteria; (3) upgrades to the
`Batch Mode' to support larger result sets with output content and formats
that can be configured by users for input into data mining applications;
(4) integration of the `By Parameter' service on the Web interface with the
Batch Mode to support queries that run too long to maintain connectivity
with a Web browser; (5) production of an XML server mode, with the many
benefits summarized above (Section~\ref{subsec:xml}) to support people
who want to write software to analyze NED data or synthesize the query
results into new VO interfaces and services.

\section{NED AS A RESOURCE FOR KDD}

Knowledge discovery in databases (KDD) refers to the complex process
of applying data mining (e.g., pattern finding) and modern statistical
analysis techniques to extract knowledge from large databases and image
archives. The first wave of results in a number of areas of scientific
research and business has made it increasingly clear that in order
to discover something fundamentally new, and to adequately handle
encounters with various \lq\lq mine fields,\rq\rq\ it is essential to
incorporate prior domain knowledge into the KDD process\cite{Fayyad96}.
As outlined above, new and emerging capabilities of NED provide a valuable
resource for incorporating prior knowledge into future KDD applications in
astronomy. For example, using the present HTML output or the future XML
server, one could write a client program that extracts information from
NED's panchromatic SEDs to fold into automated algorithms that search
for different known types of extragalactic objects in new survey data.
Likewise, a thorough comparison of new observations with images, SEDs,
high resolution spectra, or literature resources available in NED (or
distributed archive entries linked with NED by object names and
coordinates) can prevent pitfalls such as a false claim of a new class
of extragalactic object. As a third example, the NED image archive
presents an unprecedented resource for developing and
testing advanced algorithms to tackle the differing resolutions, pixel
scales, and calibration uncertainties in multi-wavelength images that
must be confronted to make novel discoveries. NED's image archive
already provides hundreds of objects that have images spanning
ultraviolet, visual, near-infrared, and radio wavelengths. In addition,
construction of intelligent Web `agent' programs that can transverse 
the {\bf External Links} from NED object queries and extract relevant 
information in an automated fashion could expand the possibilities for 
discovery in innumerable ways. 


\subsection{Fusion and Classification Using Large Astronomical Databases}
\label{subsec:fusion}

Since NED currently serves a large fraction of the world-wide extragalactic
research community with fused, multi-wavelength data for millions of
objects, it provides a unique opportunity to introduce many astronomers
to new analysis tools and protocols developed by the VO community. The
planned upgrades discussed above will enable the use of NED data streams
containing multi-wavelength, multi-dimensional data such as SEDs and
object classifications (with pointers to additional, distributed data)
in extragalactic data mining applications. NED can effectively serve this
role because the database contains 10-50 attributes (positions, redshifts,
multi-wavelength photometric measurements, and object classifications)
for millions of extragalactic objects.  Using NED as a test-bed for
data mining algorithms is a logical initial step for more ambitious VO
efforts planning to eventually handle hundreds of attributes in catalogs
containing $10^8 - 10^9$ objects or more, which is common when all types
of astrophysical sources are blended together. (Most survey catalogs
initially contain stars, Galactic nebulae, galaxies, QSOs, asteroids,
etc., until they are classified and extracted into specialized lists.)

High dimensionality presents great challenges to effectively
visualize, summarize, and extract new information from large databases.
Data fusion across large databases is fraught with practical problems:
compiling a complete sample across many wavelengths is difficult; source
cross-identifications are non-trivial to get right; there are problems of
duplicate observations, contamination, confusion, etc.; different coding
schemes for missing data must be made uniform; effects of sampling biases
must be considered; mixed data types include quantitative (continuous),
categorical (nominal), and binary (e.g., quality flags, codes); ignoring or
treating non-detections, upper-limits and different flux limits (censored
data) in combined surveys can lead to biases; observed scatter can be
intrinsic to astrophysical sources, measurement uncertainties, or both;
most classical multivariate statistical methods do not handle explicit
data uncertainties. Nevertheless, if we are to live up to the challenges
of making discoveries from fused VO archives, these and the problems of
scale will have to be confronted and solved.

\subsection{A Pilot Study}
\label{subsec:pilot}

Here we briefly summarize an ongoing project to exploit NED and its
interconnected archive resources to aid the process of automated,
large-scale galaxy classification.  A representative problem is that among
approximately a half-million objects in the Second Incremental Release
of the 2MASS Extended Source Catalog, only about 10\% were previously
known objects (established using NED, prior to loading the 2MASS sources
into the database), and only a small number of known NED objects with
2MASS cross-identifications have available morphological or spectral
classifications from the historical catalogs and literature.  A common
goal of many \lq\lq data miners\rq\rq\ in astronomy is to discover large
numbers of new cases of previously known types of objects, to perform
statistical analyses which typically suffer from small number statistics
and severe selection biases in previous investigations. Another \lq\lq
Holy Grail\rq\rq\ is the potential to discover a new class of objects,
those rare nuggets that may teach us something fundamentally new about the
contents, structure or evolution of the Universe.  Clearly using classical
approaches, even one involving manual, interactive queries of NED and
connected online archives to classify the all the previously known objects
in 2MASS (or the SDSS) is impractical. A more automated approach is needed.
An outline of the steps required for this pilot project are as follows:
(1) construct cross-identifications between near-infrared sources in 2MASS,
visual sources in SDSS, and radio sources in the NVSS and FIRST surveys;
(2) fuse the cross-correlated survey data with source classifications and
other available data in NED -- morphological types, nuclear activity types
(starburst/HII, LINER, Sy2, Sy 1, QSO, etc.), magnitudes, redshifts; (3)
comprehensively summarize correlations, dominant variables, and clusters in
the resulting high dimensionality ($N > 20$), multi-wavelength data matrix;
(4) produce a training set for machine learning classifiers (decision trees
and neural nets); (5) derive provisional classifications (predictors)
for the sources that lack any historical data. These results will guide
follow-up observations by providing candidate lists of known classes of
extragalactic objects, candidates for possible previously unknown classes
of objects, and rare objects (outliers) revealed in the multivariate
analysis. The results will be published in summary form and be made
available in bulk on the Internet as a resource for other investigators.

\section{SUMMARY}
\label{sec:summary}

NED provides data and relationships between multi-wavelength observations
of millions of extragalactic objects.  The goal is a complete panchromatic
census of objects in the extragalactic sky. NED team activities revolve
around an ongoing fusion of data from sky surveys and the literature,
focusing on established and candidate extragalactic sources, and
maintaining cross-identifications, statistical associations, and anchors
to online references and pointers. NED serves as an interface into its own
database and now also as a portal for the extragalactic research community
into an emerging federation of astrophysics data centers and service
providers with queries indexed by object names and coordinates synthesized
by NED. As a key participant in the global Virtual Observatory (VO),
NED will continue evolving the core functions for knowledge management
that it provides today, while supporting new initiatives through use of
XML standards to establish higher degrees of interoperability with other
services, cooperation in the development and application of new tools for
bulk dynamic catalog cross-identifications, serving large multidimensional
data streams from NED to interface with data mining and visualization
applications, and in general enabling new opportunities for discovery by
leveraging new information technologies.  Opportunities and challenges
to integrate NED into new VO applications have been reviewed, and we
discussed a pilot study to utilize prior knowledge on a large scale to
assemble large samples of candidates for established and potentially new
classes of objects from sky surveys. The experience of over a decade of
successful NED services and user support, combined with new capabilities
to be developed through collaboration with other organizations taking
part in construction of a global virtual observatory, is ushering in an
exciting new era of discovery using the rapidly growing online archives.

\acknowledgments

The NED and LEVEL5 development team consists of Kay Baker, Judy Bennett,
Harold Corwin, Cren Frayer, George Helou, Anne Kelly, Cheryl Lague, Joseph
Mazzarella, Barry Madore, Olga Pevunova, Marion Schmitz, Brian Skiff, and
Dianna Schettini.  The Apache Web servers are maintained by Rick Ebert,
and Informix database administration is performed by Nian-Ming Chiu. We
also thank the IPAC Systems Group for their fine computer and network
infrastructure support.  This work was carried out by the Jet Propulsion
Laboratory, California Institute of Technology, under contract with the
National Aeronautics and Space Administration (NASA). NED is funded by
the Mission Operations and Data Analysis program of NASA's Office of
Space Science. LEVEL5 is funded through the NASA Applied Information
Systems Research Program (AISRP).


\bibliography{NEDspieAstroph} 

\begin{thebibliography}{1}

\bibitem{Helou91}
G.~Helou, B.~F. Madore, M.~Schmitz, M.~D. Bicay, X.~Wu, and J.~Bennett, ``The
  {NASA/IPAC} {E}xtragalactic {D}atabase,'' in {\em Databases and Online
  Astronomy},  M.~A. Albrecht and D.~Egret, eds., pp.~89--106, Kluwer, 1991.

\bibitem{Mazz86}
J.~M. Mazzarella and V.~A. Balzano, ``A {C}atalog of {M}arkarian {G}alaxies,''
  {\em Astrophysical Journal Supplement Series{,}} {\bf 62}, pp.~751--819,
  1986.

\bibitem{Sch98}
D.~J. Schlegel, D.~P. Finkbeiner, and M.~Davis, ``{M}aps of {D}ust {I}nfrared
  {E}mission for {U}se in {E}stimation of {R}eddening and {C}osmic {M}icrowave
  {B}ackground {R}adiation {F}oregrounds,'' {\em Astrophysical Journal{,}} {\bf
  500}, pp.~525--555, 1998.

\bibitem{Bur82}
D.~Burstein and C.~Heiles, ``{R}eddenings {D}erived {F}rom {H I} and {G}alaxy
  {C}ounts - {A}ccuracy and {M}aps,'' {\em Astronomical Journal{,}} {\bf 87},
  pp.~1165--1189, 1982.

\bibitem{SciNews2001}
R.~Cowen, ``{M}ining the {S}ky,'' {\em Science News{,}} {\bf 159},
  pp.~124--125, 2001.

\bibitem{Weiss91}
J.~Weiss and J.~Good, ``{T}he {NASA} {A}strophysics {D}ata {S}ystem,'' in {\em
  Databases and Online Astronomy},  M.~A. Albrecht and D.~Egret, eds.,
  pp.~139--150, Kluwer, 1991.

\bibitem{Rut2000}
R.~Rutledge, R.~Brunner, T.~Prince, and C.~Lonsdale, ``{XID}:
  {C}ross-{A}ssociation of {ROSAT/Bright} {S}ource {C}atalog {X}-{R}ay
  {S}ources with {USNO A-2} {O}ptical {P}oint {S}ources,'' {\em Astrophysical
  Journal Supplement Series{,}} {\bf 131}, pp.~335--353, 2000.

\bibitem{Fayyad96}
U.~M. Fayyad, G.~Piatetsky-Shapiro, and P.~Smyth, ``{F}rom {D}ata {M}ining to
  {K}nowledge {D}iscovery: {A}n {O}verview,'' in {\em Advances in Knowledge
  Discovery and Data Mining},  U.~M. Fayyad, G.~Piatetsky-Shapiro, P.~Smyth,
  and R.~Uthurusamy, eds., pp.~1--34, AAII Press / The MIT Press, 1996.

\end{thebibliography}
\bibliographystyle{spiebib}   

\end{document}